\documentclass{WileyChemistry-template}
\usepackage{xcolor,cancel}
\usepackage{soul,xcolor}
\setstcolor{red}
\usepackage{bm}
\usepackage{amsmath}

\title{\raggedright Spin-filtering in a $p$-orbital helical atomic chain}

\author{
\begin{minipage}{\textwidth}
	Yasuhiro Utsumi,\textsuperscript{[a]} Takemitsu Kato,\textsuperscript{[a]} Ora Entin-Wohlman,\textsuperscript{[b]} Amnon Aharony \textsuperscript{[b]}
\end{minipage}
}

\newcommand{\affiliation}{
\begin{itemize}

\item[{[a]}] Yasuhiro Utsumi, Takemitsu Kato\\
Department of Physics Engineering, Faculty of Engineering, Mie University, Tsu, Mie, 514-8507, Japan \\
E-mail: utsumi@phen.mie-u.ac.jp

\item[{[b]}] Ora Entin-Wohlman, Amnon Aharony \\
School of Physics and Astronomy, Tel Aviv University, Tel Aviv 6997801 Israel\\

\end{itemize}
}

\renewcommand{\abstract}{
We theoretically analyze spin filtering in two-terminal systems,  induced by the spin-orbit interaction (SOI), as a possible origin of the ``chirality-induced spin selectivity'' (CISS) effect observed experimentally in chiral molecules, such as DNA.
Due to Bardarson's theorem, spin filtering cannot be realized in a molecule containing one orbital-channel.
However,  when two orbitals are involved, SOI can induce spin filtering in a molecule coupled to two terminals without braking time-reversal symmetry.
In particular, we provide an example of a $4 \times 4$ reflection matrix for a spinful electron passing through a molecule containing two orbital-channels, which complies with Bardarson's theorem and produces a finite spin conductance.
As a microscopic toy model realizing a single strand of DNA, we consider a $p$-orbital helical atomic chain with intra-atomic SOI's and a strong crystalline field along the helix.  
This model exhibits two-orbital spin filtering:
For various parameters preserving the helical symmetry, the model hosts spin asymmetric states carrying pairs of up and down spins propagating in opposite directions.
 The typical energy scale of the helical states is the product of the intra-atomic SOI and the curvature. The large value of this energy identifies our model as a likely candidate to explain the CISS in organic molecules. 
}
\newcommand{\keywords}{
	Chirality induced spin selectivity effect \textbullet\
	Spin orbit interaction \textbullet\
	Helical symmetry \textbullet\
	Time reversal symmetry
}

\begin{document}

\twocolumn[\vspace{-1.5cm}\maketitle\vspace{-1cm}
	\textit{\dedication}\vspace{0.4cm}]
\small{\begin{shaded}
		\noindent\abstract
	\end{shaded}
}

\begin{figure} [!b]
\begin{minipage}[t]{\columnwidth}{\rule{\columnwidth}{1pt}\footnotesize{\textsf{\affiliation}}}\end{minipage}
\end{figure}




\section*{Introduction}
\label{introduction}

Electrons injected into chiral molecules like DNA become spin polarized after being transmitted through the molecule.~\cite{Goehler2011,Xie2011,Mishra2020,Naaman2019,Waldeck2021}.
Such a spin-filtering effect has been termed ``chirality-induced spin selectivity" (CISS)~\cite{Naaman2012,Michaeli2016,Michaeli2017}.
This is a remarkable effect, since organic molecules do not contain magnetic atoms, which would be apparent candidates for spin-dependent phenomena.

Although CISS is established experimentally, its theoretical understanding is still debated~\cite{Evers2022}.
In many theories, the spin-orbit interaction (SOI) is considered to be the origin of the spin asymmetry~\cite{Guo2012,Gutierrez2013,Guo2014,Matityahu2016,Michaeli2019,Geyer2020,YU2020,SierraBioMol2020,Liu2021,Wolf2022,Michaeli2022}.
However, since the SOI does not break time-reversal symmetry, the appearance of  SOI-induced spin-filtering is a non-trivial effect:
Bardarson's theorem \cite{Bardarson2008} imposes a constraint stating that \textit{in time-reversal-symmetric systems with half integer spins, the transmission eigenvalues of the scattering matrix come in degenerate pairs}.
Assuming that this Kramers-type degeneracy carries up and down spins in the same direction, the theorem prohibits spin filtering in systems coupled to two terminals.
However, the theorem forbids spin selectivity through two-terminal time-reversal symmetric systems only when there is only one orbital-channel.
Therefore, several previous theories broke time-reversal symmetry by introducing spin dissipation~\cite{Guo2012,Guo2014,Matityahu2016}, which effectively introduces many terminals.

Another option to overcome Bardarson's theorem, recently formulated explicitly~\cite{YU2020}, is to introduce two-orbital conducting channels.
Bardarson's theorem does not specify which spin states are associated with the doubly-degenerate transmission eigenvalues.
Therefore, spin-filtering is possible if there exist two pairs of doubly-degenerate transmission eigenvalues, in which one pair carries two up spins in one direction and the other pair carries two down spins in the opposite direction.
The origin of this idea is spin filtering brought about by the Rashba SOI in two-terminal quantum point contacts~\cite{Eto2005}, in tubular two-dimensional gases~\cite{Entin-WohlmanPRB2010}, and in quasi-one dimensional quantum wires~\cite{NagaevPRB2014}.
In the context of CISS, the idea appeared implicitly in the models of a particle traveling on~\cite{Michaeli2019,Geyer2020} or along~\cite{Gutierrez2013} the surface of a helical tube, and in the double-helix model with two orbitals residing on different helices~\cite{SierraBioMol2020}.

In a previous paper~\cite{YU2020}, we demonstrated that a $p$-orbital helical atomic chain, a toy model of a single strand of the DNA molecule, can be reduced to an effective two-orbital tight-binding model realizing a two-terminal spin filter  for specific parameters, without breaking time-reversal symmetry.
In that paper we mainly focused on an ideal configuration: two pairs of up and down spins propagating in opposite directions.
In the present paper we extend our previous work to a broader range of parameters.
Especially, we analyze the helical symmetry of our model for all $p$-orbital states.
The consequences of including the orbital degrees of freedom has been discussed before:
(i) Orbital polarization emerges in the $p$-orbital helical atomic chain~\cite{Otsuto2021};  (ii) The sign of the  hopping matrix elements in a neighboring $p$-$p$ block is related to the chirality and to the direction of spin-polarization~\cite{Zoellner2020}.
In particular we discuss the consequences of  the helical symmetry of the hopping matrix elements connecting neighboring $p$-orbitals.
Since the DNA molecule is complex and it is difficult to derive systematically its effective $p$-orbital tight-binding model, we only account for the constraint imposed by the helical symmetry.
In the following, we use $\hbar=1$.


\section*{Spin filtering in a two-terminal two-orbital time-reversal symmetric conductor}
\label{sec:ttsf}

\begin{figure}
\begin{center}
\includegraphics[width=1.0 \columnwidth]{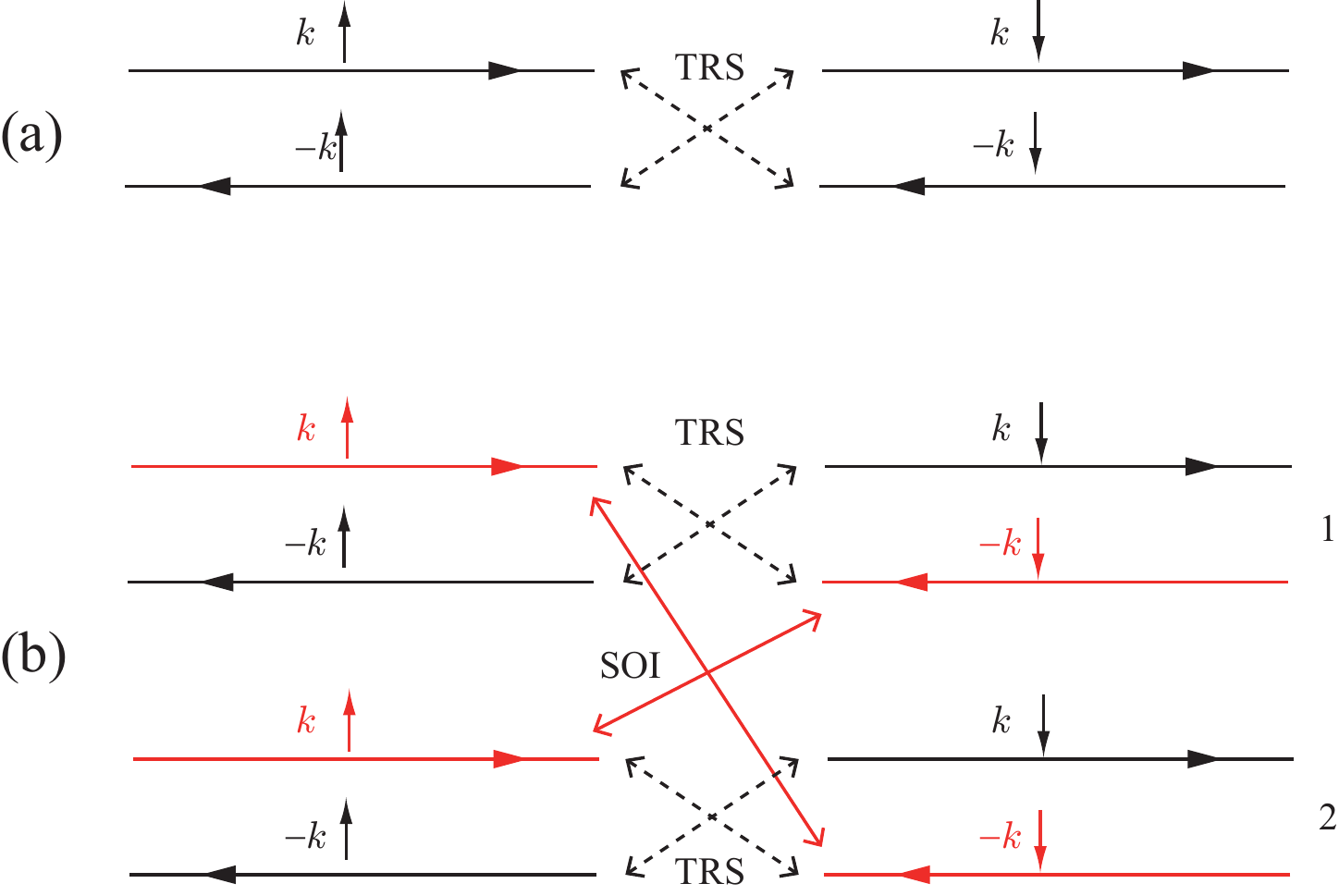}
\caption{
(a) Single-orbital case.
For each spin, there are left- and right-going states.
Time-reversed states (TRSs) are indicated by dashed arrows.
The Hamiltonian, which mixes up and down spins propagating in the opposite directions is odd under the time-reversal operation.
(b) Two-orbital case.
The Hamiltonian, which mixes states connected by solid arrows can be even under the time-reversal operation. }
\label{fig:even_helical}
\end{center}
\end{figure}

We begin by briefly explaining the reason why the SOI, which does not break time-reversal symmetry, cannot realize two-terminal spin filtering in a single-orbital conducting channel.
In such a system, there are 2 channels when the spin degree of freedom is accounted for [Fig.~\ref{fig:even_helical} (a)]:
For each spin $\sigma$, there are right- and left-going modes, $|k; \sigma \rangle$ and $|-k; \sigma \rangle$.
Spin filtering occurs if, e.g., the right-going $\downarrow$-spin and left-going $\uparrow$-spin states, $|k;\downarrow \rangle$ and $|-k;\uparrow \rangle$, dominate the transport.
For this purpose, the other two states, $|k;\uparrow \rangle$ and $|-k;\downarrow \rangle$, have to be gapped away.
Note that they are time-reversed of one another, $\hat{\Theta}|k;\sigma \rangle = \sigma |-k; \bar{\sigma} \rangle$, where $\hat{\Theta}=-i \hat{\sigma}_y \hat{K}$  is the time-reversal operator ($\hat{\sigma}_y$  is the Pauli matrix and $\hat{K}$ is the complex conjugate operator)~\cite{JJSakurai1985}.
Here $\bar{\sigma}=\downarrow (\uparrow)$ for $\sigma=\uparrow (\downarrow)$
and the coefficient $\sigma$ has the values $\sigma = +1 (-1)$ for $\sigma=\uparrow (\downarrow)$.
The Hamiltonian which hybridizes time-reversed states is given by,
\begin{align}
\hat{V} = a \hat{c}_{k;\uparrow}^\dagger \hat{c}^{}_{-k;\downarrow} + a^* \hat{c}_{-k;\downarrow}^\dagger \hat{c}^{}_{k;\uparrow} \, , \label{eqn:V}
\end{align}
where $a$ is a complex number. 
Since the annihilation operator transforms as $\hat{\Theta} \hat{c}^{}_{k;\sigma} \hat{\Theta}^{-1} = \sigma \hat{c}^{}_{-k; \bar{\sigma}}$ ~\cite{Bernevig2013},
the mixing Hamiltonian is odd under the time reversal operation as $\hat{\Theta} \hat{V} \hat{\Theta}^{-1} = - \hat{V}$.
Consequently, it is not possible to realize spin filtering without breaking time-reversal symmetry in a single-orbital conducting channel.

The situation is different when there are two orbital channels, which we denote as orbital $1$ and orbital $2$ [Fig.~\ref{fig:even_helical} (b)].
In this case one may consider hybridizing the right-going $\uparrow$-spin and the left-going $\downarrow$-spin residing on different orbitals. 
Such a Hamiltonian,
\begin{align}
\hat{V}^\prime =& a \, \hat{c}_{k;2, \uparrow}^\dagger \hat{c}^{}_{-k;1, \downarrow} - a \, \hat{c}_{k;1, \uparrow}^\dagger \hat{c}^{}_{-k;2, \downarrow} \nonumber \\
&+ a^* \hat{c}_{-k;1, \downarrow}^\dagger \hat{c}^{}_{k;2, \uparrow} - a^* \hat{c}_{-k;2, \downarrow}^\dagger \hat{c}^{}_{k;1, \uparrow} \, , \label{eqn:Vp}
\end{align}
 where $a$ is a complex number, is even under the time-reversal operation, $\hat{\Theta} \hat{V}^\prime \hat{\Theta}^{-1} = \hat{V}^\prime$.
Therefore, the SOI can in principle lead to spin filtering when there are two-orbital channels.

In the next section we provide a different argument, based on the scattering matrix of the molecule, and demonstrate that spin filtering by a two-terminal two-orbital  setup  does not contradict Bardarson's theorem~\cite{Bardarson2008}.

\subsection*{Scattering matrix}
\label{sec:Bardarson_T}

Let us consider a chiral molecule attached to left and right leads.
The scattering states in the left ($s=L$) and right ($s=R$) leads are,
\begin{align}
|\psi \rangle^{}_s = |{k} \rangle^{}_s c_s^{\mathrm  in}  + |-{k} \rangle^{}_s c_s^{\mathrm  out}  \, .
\label{eqn:scattering_state_}
\end{align}
The scattering matrix ${S}$, which connects the amplitude of right-going $|{k} \rangle^{}_s$ and left-going $|-{k} \rangle^{}_s$ states,
\begin{align}
{c}^{\mathrm  out} = {S} c^{\mathrm  in} \, ,
\;\;\;\;
{c}^{\mathrm  out} = \left[ \begin{array}{c} {c}_L^{\mathrm  out} \\ {c}_R^{\mathrm  out} \end{array} \right] \, ,
\;\;\;\;
c^{\mathrm  in}= \left[\begin{array}{c} c_L^{\mathrm  in} \\ c_R^{\mathrm  in} \end{array} \right]\, ,
\label{eqn:s_def_1}
\end{align}
reads
\begin{align}
{S} = \left[\begin{array}{cc} {r} & {t}' \\ {t} & {r}' \end{array} \right] \, .
\label{eqn:s_matrix}
\end{align}

When there are $N^{}_s$ orbital channels in terminal $s$, the amplitudes $c^{\mathrm in}_{s}$ and ${c}^{\mathrm out}_{s}$ form 2$N^{}_s-$component vectors,
\begin{align}
c_s^{{\mathrm in}} = \left[ \begin{array}{c} c_{1 \uparrow s}^{{\mathrm in} } \\ c_{1 \downarrow s}^{{\mathrm in} } \\ \vdots \\ c_{N_s \uparrow s}^{{\mathrm in} } \\ c_{N_s \downarrow s}^{{\mathrm in} }
\end{array} \right]\ ,
\;\;\;\;
{c}_s^{{\mathrm out}} = \left[ \begin{array}{c} {c}_{1 \downarrow s}^{{\mathrm out} } \\ {c}_{1 \uparrow s}^{{\mathrm out}} \\ \vdots \\ {c}_{N_s \downarrow s}^{{\mathrm out} } \\ {c}_{N_s \uparrow s}^{{\mathrm out}} \end{array} \right] \, .
\label{amp}
\end{align}
For a time-reversal symmetric system, the scattering matrix is self-dual~\cite{Bardarson2008},
\begin{align}
S=({\bm 1}_{2 N^{}_{s}} \otimes \sigma^{}_y) S^T ({\bm 1}_{2 N^{}_{s}} \otimes \sigma^{}_y) \ ,
\label{eqn:BT}
\end{align}
where $S^{T}_{}$ is the transposed scattering matrix (${\bm 1}_{ n}$ is the $n \times n$ unit matrix).
The block-diagonal component of the scattering matrix is the reflection matrix, $r=({\bm 1}_{N^{}_{s}} \otimes \sigma^{}_y) r^T ({\bm 1}_{N^{}_{s}} \otimes \sigma^{}_y)$.
Hence, the reflection amplitude from the state with orbital $\alpha'$ and spin $\sigma'$ into the state with orbital $\alpha$ and spin $\sigma$, $r^{}_{\alpha \sigma,\alpha' \sigma'}$, satisfies,
\begin{align}
r^{}_{\alpha \sigma , \alpha' \sigma'} = \sigma \sigma' \, r^{}_{\alpha' \bar{\sigma}' , \alpha \bar{\sigma}} \, . \label{eqn:sym_ref}
\end{align}
The transmission eigenvalues mentioned above are the eigenvalues of the matrix of transmission probabilities,
\begin{align}
{t}^\dagger {t} = {\bm 1}_{2 N_s}- r^\dagger_{} r  \, .
\end{align}
For the single-orbital channel, $N_{s}^{}=1$, the reflection matrix is a $2 \times2$ matrix.
It is diagonal \cite{Bardarson2008}, as by Eq. (\ref{eqn:sym_ref})
\begin{align}
r^{}_{\uparrow , \uparrow} =& r^{}_{\downarrow , \downarrow}=r^{}_0 , \\
r^{}_{\sigma , \bar{\sigma}} =& -r^{}_{\sigma , \bar{\sigma}}=0.
\end{align}
The matrix of transmission probabilities is also diagonal, ${t}^\dagger {t} = (1-|r^{}_0|^2) \sigma^{}_0$.
Therefore, the transmission eigenvalues are degenerate.
Since spin asymmetry is absent, spin filtering is forbidden.

For the two-orbital channel case, $N_{s}^{}=2$, the reflection matrix is a $4 \times 4$ matrix.
A simple example, which satisfies (\ref{eqn:sym_ref}) and is capable of producing spin filtering, is~\cite{YU2020}:
\begin{align}
r &= \left[ \begin{array}{cccc}
0 & 0^{}_{} & 0^{}_{} & r^{}_{1 \uparrow , 2 \downarrow} \\
0 & 0 & r^{}_{1 \downarrow , 2 \uparrow} & 0 \\
0 & r^{}_{2 \uparrow, 1 \downarrow} & 0 & 0 \\
r^{}_{2 \downarrow , 1 \uparrow} & 0 & 0 & 0
\end{array} \right]\nonumber \\
&= \left[ \begin{array}{cccc}
0 & 0^{}_{} & 0^{}_{} & r^{}_{1 \uparrow , 2 \downarrow} \\
0 & 0 & -r^{}_{2 \downarrow , 1 \uparrow} & 0 \\
0 & -r^{}_{1\uparrow,2\downarrow} & 0 & 0 \\
r^{}_{2 \downarrow , 1 \uparrow} & 0 & 0 & 0
\end{array} \right] \ .
\label{eqn:rm_2c}
\end{align}
The matrix (\ref{eqn:rm_2c}) can be rearranged in a block-diagonal form,
$r=\mathrm{diag}(r^{}_+,r^{}_-)$, where
\begin{align}
r^{}_+ = \left[ \begin{array}{cc} 0 & r^{}_{1 \uparrow , 2 \downarrow} \\ r^{}_{2 \downarrow , 1 \uparrow} & 0 \end{array} \right] ,\
r^{}_- = \left[ \begin{array}{cc} 0 & -r^{}_{1 \uparrow , 2 \downarrow} \\ -r^{}_{2 \downarrow , 1 \uparrow} & 0 \end{array} \right] \ . \label{eqn:rp_rm}
\end{align}
The two matrices  $r^{}_{+}$ and $r^{}_{-}$ are time-reversed of one another,  $r^{}_- = \sigma^{}_y r_+^T \sigma^{}_y$.
The four transmission eigenvalues are the solutions of the characteristic polynomial equation
\begin{align}
\mathrm{det} \{ \Lambda {\bm 1}_{4}- t^\dagger t \} &= ( \mathrm{det} \{ (\Lambda-1) {\bm 1}^{}_{2} + r_\pm^\dagger r^{}_\pm \})^2=0 \ .
\end{align}
They come in pairs of degenerate eigenvalues~\cite{Bardarson2008},
\begin{align}
1+|r_{1 \uparrow, 2 \downarrow}|^2, 1+|r_{1 \uparrow, 2 \downarrow}|^2, 1+|r_{2 \downarrow, 1 \uparrow}|^2, 1+|r_{2 \downarrow, 1 \uparrow}|^2 \, .
\end{align}

Let us examine the spin filtering associated with the reflection matrix (\ref{eqn:rm_2c}).
The spin conductance at the left lead is given by~\cite{YU2020},
\begin{align}
G^{}_{j;LL} =  {\mathrm Tr} [ {\bm 1}^{}_2 \otimes \sigma^{}_j ( {\bm 1}^{}_4 - r r^\dagger _{}) ]/(2\pi)\ , \label{eqn:Gj}
\end{align}
where $\sigma^{}_{j}$ ($j=x,y,z$) is the $j$th Pauli matrix.
Inserting Eq. (\ref{eqn:rm_2c}) into Eqs.~(\ref{eqn:Gj}) yields a finite spin conductance for the $z$ component of the spin
\begin{align}
G^{}_{z;LL} =( |r^{}_{2 \downarrow, 1 \uparrow}|^2 - |r^{}_{1 \uparrow, 2 \downarrow}|^2 )/\pi \ .
\end{align}
The spin conductances for the other components vanish, $G^{}_{x;LL} =G^{}_{y;LL}=0$.
The charge conductance is obtained from Eq.~(\ref{eqn:Gj}) by replacing the Pauli matrix with the unit matrix $\sigma_0={\bm 1}_2$,
\begin{align}
G^{}_{0;LL} =(2 -  |r^{}_{2 \downarrow, 1 \uparrow}|^2 - |r^{}_{1 \uparrow, 2 \downarrow}|^2)/\pi\ ,
\end{align}
It follows that the (normalized) spin polarization is
\begin{align}
P^{}_{z;L} = \frac{G^{}_{z;LL}}{G^{}_{0;LL}} = \frac{ |r_{2 \downarrow, 1 \uparrow}|^2 - |r_{1 \uparrow, 2 \downarrow}|^2 }{2 -  |r^{}_{2 \downarrow, 1 \uparrow}|^2 - |r^{}_{1 \uparrow, 2 \downarrow}|^2} \ .
\label{eqn:pzL}
\end{align}
Perfect spin-filtering $P^{}_{z;L} =1$ (or $P^{}_{z;L} =-1$) is achieved for
$|r_{2 \downarrow, 1 \uparrow}|^2=1$ (or $|r^{}_{1 \uparrow, 2 \downarrow}|^2=1$). The condition $|r_{2 \downarrow, 1 \uparrow}|^2=1$ corresponds to the Hamiltonian (\ref{eqn:Vp}).

The specific reflection matrix in the example (\ref{eqn:rm_2c}) demonstrates the possibility of spin filtering in a two-terminal molecule.
However, it is still a non-trivial task to construct a microscopic model realizing the two-orbital spin-filter.
In the next section, following our previous work~\cite{YU2020}, we describe a $p$-orbital helical atomic chain realizing the spin-filtering when it is attached to two terminals.

\section*{A $p$-orbital helical tight-binding model}
\label{sec:effective_hamiltonian}

\begin{figure}[ht]
\begin{center}
\includegraphics[width=1 \columnwidth]{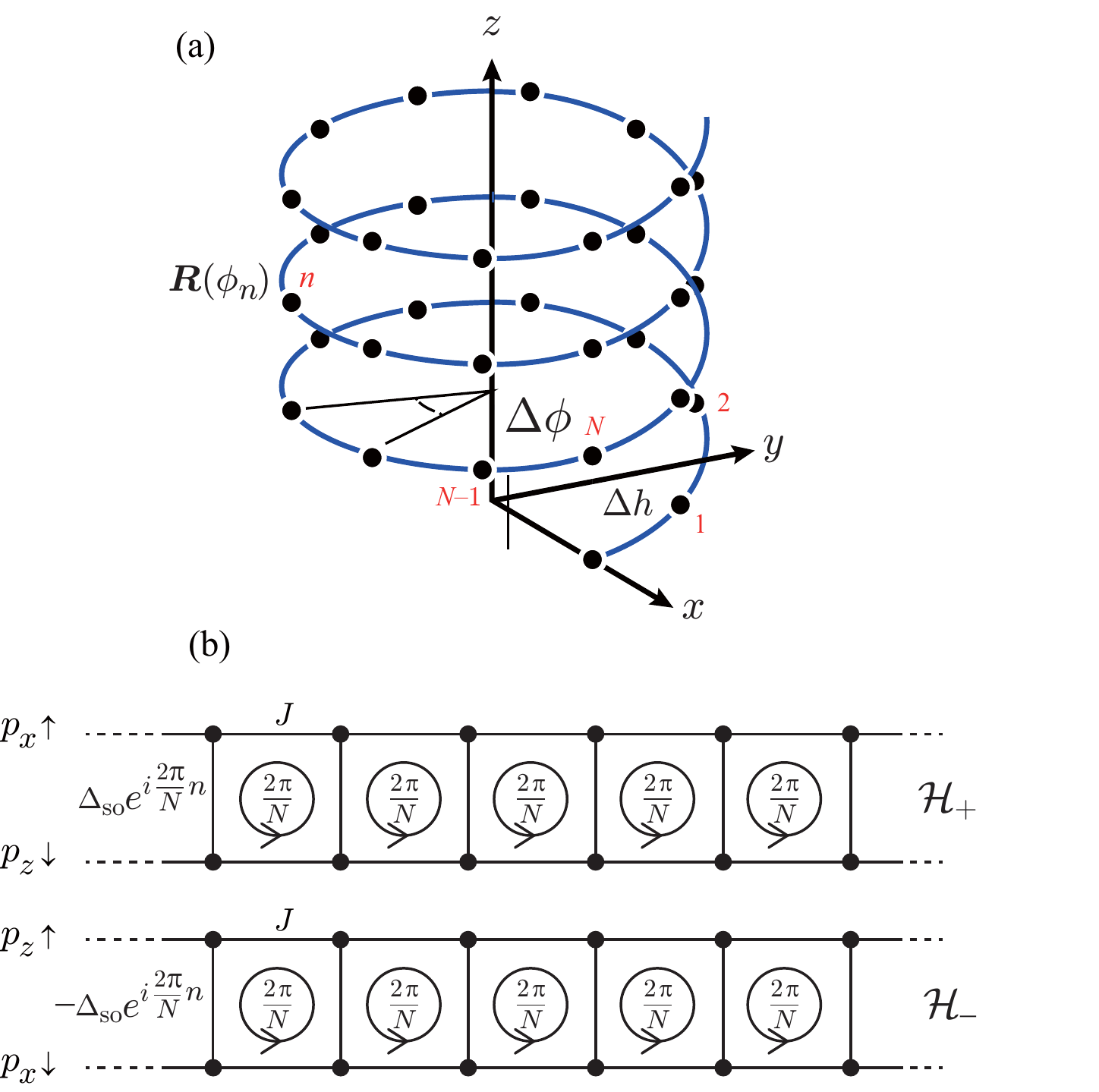}
\caption{
(a) Schematic picture of a $p$-orbital helical atomic chain, a toy model for a single strand of a double-stranded DNA.
${\bm R}(\phi^{}_{n})$ is the radius-vector to the $n$th site, within the Frenet-Serret frame [Eq. (\ref{Rphi})],
$\Delta h$ is the pitch, $\Delta\phi=2\pi/N$, and $\phi^{}_{n}=n p \Delta\phi$.
(b) Ladders threaded by a fractional flux $2\pi/N$.
The vertical lines represent the tunneling amplitudes
$\pm p \Delta_{so} \exp(ip \phi_n)$ connecting $\uparrow$- and $\downarrow$-spins on different orbitals at the $n$th rung.
The site index $n$ increases from left to right.
}
\label{fig:setupDNA}
\end{center}
\end{figure}

Here we summarize the construction of the tight-binding Hamiltonian describing the $p$-orbital helical atomic chain shown in Fig.~\ref{fig:setupDNA} (a) (see Appendix B of Ref.~\cite{YU2020}). 
The vector from the origin to a point on a continuous helix of radius $R$ and pitch $\Delta h$ is, 
\begin{align}
{\bm R}(\phi) = \left [  R \cos ( \phi),R \sin (p \phi), \Delta h \, \phi/(2 \pi) \right] \ ,
\label{Rphi}
\end{align}
where $p=1$ ($p=-1$) for a helix twisted in the right-handed (left-handed) sense.
In the Frenet-Serret frame, the tangent ${\bm t}$ (along  the helix), normal ${\bm n}$, and bi-normal ${\bm b}$ unit vectors at the point on the helix are
\begin{align}
&{\bm t}(\phi) = \left [ - \kappa \sin (\phi), p \kappa \cos (\phi), |\tau| \right]\ ,\nonumber\\
&{\bm n}(\phi) = \left [- \cos (\phi), - p \sin (\phi), 0 \right] \ ,\nonumber\\
&{\bm b}(\phi) = {\bm t}(\phi) \times {\bm n}(\phi) = \left [ p |\tau| \sin (\phi), - |\tau| \cos (\phi), p \kappa \right] \ ,
\label{tnb}
\end{align}
where the `normalized' curvature and torsion, $\kappa$ and $\tau$, are
\begin{align}
\kappa = \frac{R}{\sqrt{R^2+[\Delta h/(2 \pi)]^2}} , \;\; \tau = \frac{p \Delta h/(2 \pi)}{\sqrt{R^2+[\Delta h/(2 \pi)]^2}}  .  \label{eqn:tau}
\end{align}

The position of the $n$th site in the tight-binding scheme is specified by ${\bm R}(\phi_n)$, where the increment of $\phi$ between neighboring sites is $\Delta\phi=2\pi/N$, and $\phi^{}_n = p 2 \pi n/N$.
The tight-binding Hamiltonian of the helical atomic chain is,
\begin{align}
\hat{\mathcal H}^{}_{\mathrm mol} =&\Big (\sum_{n}^{} - \hat{c}_{n+1}^\dagger {\bm J} \otimes \sigma^{}_0 \tilde{c}^{}_n  + {\mathrm H.c.} \Big ) \nonumber \\ &+ \sum_{n}^{} \epsilon^{}_0 \, \hat{c}_{n}^\dagger \hat{c}^{}_{n} - 2 \Delta^{}_{so} \, \hat{c}_n^\dagger {\bm L} \cdot {\bm S} \hat{c}^{}_n  \nonumber \\ &+ \sum_{n}^{} K_{ {\bm t} } \, \hat{c}_n^\dagger [ ( {\bm t}(\phi_n) \cdot {\bm L})^2 - {\bm 1}_3 ] \hat{c}^{}_n  \nonumber \\ &+ \Delta \epsilon \, \hat{c}_n^\dagger  [ ( {\bm b}(\phi_n) \cdot {\bm L})^2 - ( {\bm n}(\phi_n) \cdot {\bm L})^2 ] \hat{c}^{}_n \ ,
\label{eqn:original_hamiltonian}
\end{align}
 where
\begin{align}
\hat{c}^{\dagger}_{n} = \left[ \begin{array}{cccccc} \hat{c}^{\dagger}_{n;p_x \uparrow} & \hat{c}^{\dagger}_{n;p_x \downarrow} & \hat{c}^{\dagger}_{n;p_y \uparrow} & \hat{c}^{\dagger}_{n;p_y \downarrow} & \hat{c}^{\dagger}_{n;p_z \uparrow} & \hat{c}^{\dagger}_{n;p_z \downarrow} \end{array} \right] \ , \label{eqn:vec_c}
\end{align}
is the vector of creation operators: $\hat{c}_{n;o \sigma}^\dagger$ is a creation operator of an electron residing at site $n$ with orbital $o$ and spin $\sigma$.
The first term on the right-hand side of Eq.~(\ref{eqn:original_hamiltonian}) describes the hopping between nearest-neighbor sites, with the hopping amplitude ${\bm J}$ being a $3 \times 3$ matrix.
Since the system is time-reversal symmetric, all the components  of the matrix are real
\begin{align}
\hat{\Theta} {\bm J} \hat{\Theta}^{-1}  = \hat{K} {\bm J} \hat{K}^{-1} = {\bm J} ,
\end{align}
where we have used $\Theta \hat{c}^{}_{n;\alpha} \Theta^{-1} = -i  {\bm 1}_3 \otimes \sigma^{}_y \hat{c}^{}_{n;\alpha}$.
In the second term, $\epsilon_0$ is the on-site energy.
The third term represents the intra-atomic spin-orbit interaction whose strength is denoted $\Delta^{}_{so}$.
Here ${\bm L}= (L_x,L_y,L_z)$ is the vector of the orbital angular-momentum operators,
\begin{align}
L^{}_x =& \left[ \begin{array}{ccc} 0 & 0 & 0 \\ 0 & 0 & -i \\ 0 & i & 0 \end{array} \right], \\
L^{}_y =& \left[ \begin{array}{ccc} 0 & 0 & i \\ 0 & 0 & 0 \\ -i & 0 & 0 \end{array} \right], \\
L^{}_z =& \left[ \begin{array}{ccc} 0 & -i & 0 \\ i & 0 & 0 \\ 0 & 0 & 0 \end{array} \right]
,
\end{align}
and ${\bm S}={\bm \sigma}/2$ is the  vector of the spin angular-momentum, with ${\bm \sigma}$ being the vector of the Pauli matrices.

The other terms in the Hamiltonian describe the crystalline  fields created by neighboring atoms:
$K_{\bm t}$ is the orbital anisotropy energy along the tangential direction ${\bm t}(\phi_n)$.
$\Delta \epsilon$ is the difference between the orbital anisotropy energies along the normal direction ${\bm n}(\phi_n)$ and the bi-normal direction ${\bm b}(\phi_n)$.
We assume the leading orbital anisotropy energy is the one along the tangential direction, $K_{\bm t}$.
 This condition would effectively mimic the situation discussed in the helical tube models~\cite{Michaeli2019,Geyer2020,Gutierrez2013} in a simple manner.

\subsection*{Helical symmetry}
\label{Helical_symmetry}

In the following, we analyze the helical symmetry for the infinite chain. 
Although the system is finite in the transport experiment, for a sufficiently long molecule, the transport properties would be dominated by the bulk electric states. 
Then the system we consider possesses helical symmetry, i.e., the Hamiltonian is invariant under the screw operation~\cite{Otsuto2021}: a translation by one site and a rotation  by $p \Delta \phi = p 2 \pi/N$ along the $z$ axis,
\begin{equation}
\hat{D}_z(p \Delta \phi) \hat{T} \hat{\mathcal H}^{}_{\mathrm mol} \hat{T}^{-1} \hat{D}_z^{-1}(p \Delta \phi) =  \hat{\mathcal H}^{}_{\mathrm mol} .
\end{equation}
The translation operator $\hat{T}$ shifts the site index of the operators by one,
\begin{align}
\hat{T} \hat{c}_{n;o,\sigma} \hat{T}^{-1} = \hat{c}_{n+1;o,\sigma} \, .
\end{align}
The rotation operator around the $z$ axis is,
\begin{align}
\hat{D}_z(p \Delta \phi) = e^{-i (\hat{L}_z + \hat{S}_z) p \Delta \phi} ,
\end{align}
where
$\hat{L}_z=\sum_n \hat{c}_n^\dagger \left( L_z \otimes {\bm 1}_2 \right) \hat{c}_n$ and
$\hat{S}_z=\sum_n \hat{c}_n^\dagger \left( {\bm 1}_3 \otimes S_z \right) \hat{c}_n$.
Thus, the rotation changes Eq.~(\ref{eqn:vec_c}) to be
\begin{align}
\hat{D}_z(p \Delta \phi) \hat{c}_n \hat{D}_z^{-1}(p \Delta \phi) = e^{i L_z p \Delta \phi} \otimes  e^{i S_z p \Delta \phi} \hat{c}_n \, .
\end{align}

This operation does not change the on-site energy term in the Hamiltonian.
The intra-atomic SOI does not change either since $[L_z+S_z, {\bm L} \cdot {\bm S}]=0$.
The crystalline field terms in the third and forth lines of (\ref{eqn:original_hamiltonian}) do not change as well.
This can be verified by exploiting the following relations:
\begin{align}
{\bm t}(\phi_{n+1}) \cdot {\bm L} =&  e^{-i L_z p\Delta \phi} [ {\bm t}(\phi_{n}) \cdot {\bm L} ] e^{i L_z p\Delta \phi} \, , \\
{\bm n}(\phi_{n+1}) \cdot {\bm L} =&  e^{-i L_z p\Delta \phi} [ {\bm n}(\phi_{n}) \cdot {\bm L} ] e^{i L_z p\Delta \phi} \, , \\
{\bm b}(\phi_{n+1}) \cdot {\bm L} =&  e^{-i L_z p\Delta \phi} [ {\bm b}(\phi_{n}) \cdot {\bm L} ] e^{i L_z p\Delta \phi} \, .
\end{align}
The first (hopping)  term is transformed as,
\begin{align*}
\sum_{n}^{} \hat{c}_{n+1}^\dagger {\bm J} \otimes \sigma^{}_0 \hat{c}^{}_n
\to
\sum_{n}^{} \hat{c}_{n+2}^\dagger e^{-i L_z p \Delta \phi} {\bm J} e^{i L_z p \Delta \phi} \otimes \sigma^{}_0 \hat{c}^{}_{n+1} \, .
\end{align*}
Therefore, the hopping matrix ${\bm J}$ satisfies,
\begin{align}
\bm J =e^{-i L_z p \Delta \phi} {\bm J} e^{i L_z p \Delta \phi} \, . \label{eqn:constraint}
\end{align}
Consequently, the elements of the hopping matrix are parameterized by three parameters, $J$, $\alpha$ and $\varphi$ as,
\begin{align}
{\bm J} = J \left[ \begin{array}{ccc} \alpha \cos \varphi & - \alpha \sin \varphi & 0 \\ \alpha \sin \varphi  & \alpha \cos \varphi & 0 \\ 0 & 0 & 1 \end{array} \right] .
\end{align}
The three parameters are real numbers, since the Hamiltonian is time-reversal symmetric.

In our previous work~\cite{YU2020}, we demonstrated the two-terminal two-orbital spin filtering for a specific condition: $K_t \to \infty$, $\varphi=-p \Delta \phi$ and $\alpha=1$.
In the next section, we analyze the filtering for other various parameters.

\section{Band structure}
\label{sec:band_structure}

\begin{figure*}[ht]
\begin{center}
\includegraphics[width=16.4cm]{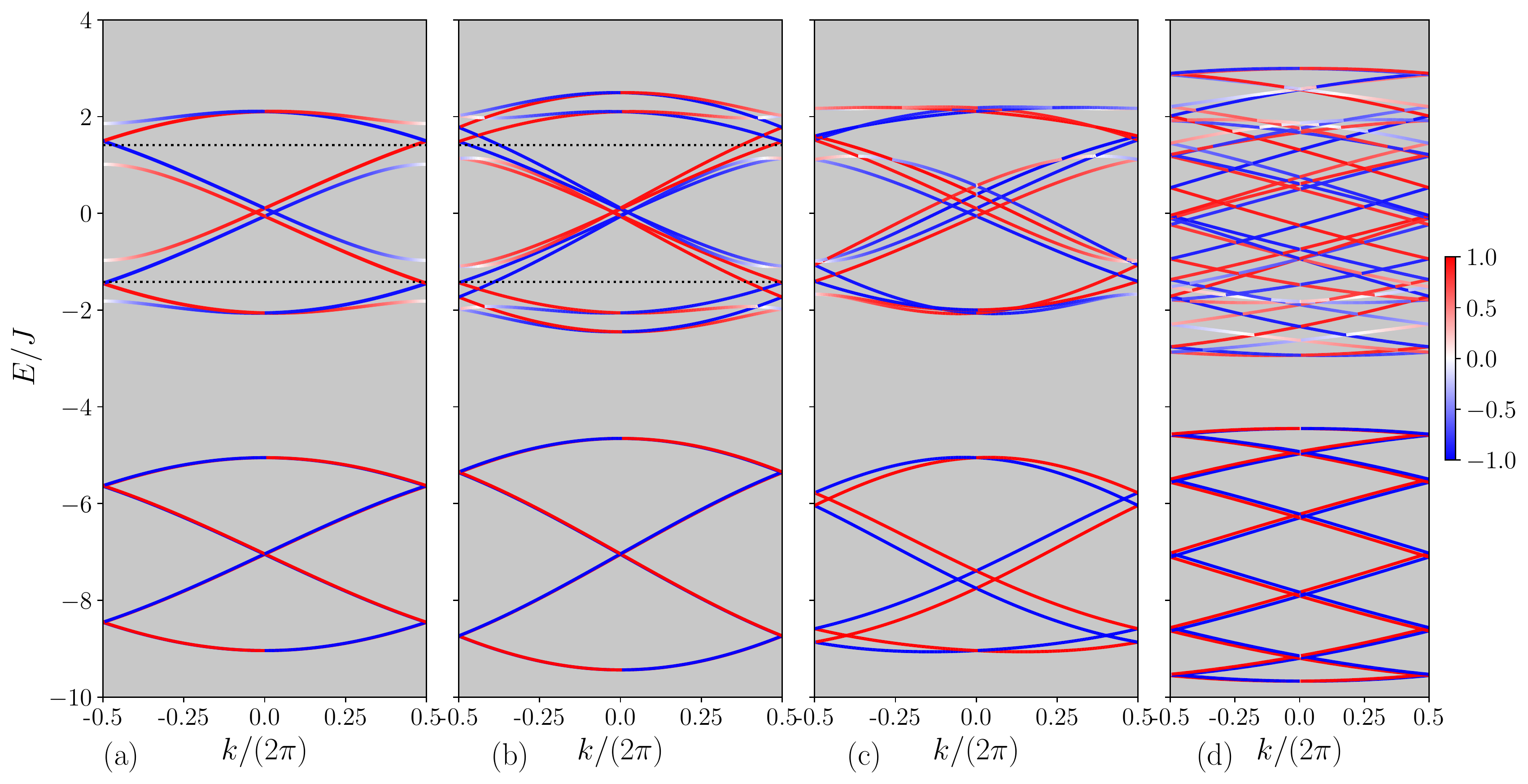}
\caption{Band structures for various parameters
:
(a) $\alpha=1$, $\varphi=-\Delta \phi$, $\tau=0$ and $N=4$,
(b) $\alpha=1.2$, $\varphi=-\Delta \phi$, $\tau=0$ and $N=4$
(c) $\alpha=\sqrt{2}$, $\varphi=\pi/4$, $\tau=0$ and $N=4$,
and
(d) $\alpha=\sqrt{2}$, $\varphi=\pi/4$, $\tau=0.48$ and $N=10$.
Other parameters are fixed as $p=1$, $\Delta_{so}=0.4 J$, $\epsilon_0=\Delta \epsilon=0$ and $K_t=7 J$. 
The color scheme indicates the $z$ component of the average spin (red for $\uparrow$ spin and blue for $\downarrow$  spin, see the color bar). 
}
\label{fig:band}
\end{center}
\end{figure*}

Figure \ref{fig:band} shows the band structure in the reduced zone scheme for various parameters, obtained by imposing periodic boundary condition $\hat{c}^{}_{MN+n}=\hat{c}^{}_n$ with $M \to \infty$.
The color scheme indicates the $z$ component of the average spin (red for $\uparrow$ spin and blue for $\downarrow$  spin, see the color bar). 
In the following, we fix $p=1$, and $\epsilon_0=\Delta \epsilon=0$.
The strength of the SOI is taken as $\Delta^{}_{so}=0.4 J$.
This estimation is based on a band of width $4J \sim 120 \mathrm{meV}$~\cite{Gutierrez2012} and the intra-atomic SOI energy in carbon nanotubes $\Delta_{so} \sim 12 \mathrm{meV}$~\cite{HuertasHernandoPRB2006}.
The crystalline field is taken to be sufficiently large, as $K_t=7J$.
Due to  this strong crystalline field along the tangential direction of the helix, $K_t$, there are two energetically split bands.
The lower band is the $\sigma$-band and the upper band is the $\pi$-band.

Panel (a) in Fig. \ref{fig:band} shows the band structure for $\alpha=1$, $\varphi=-\Delta \phi$ and $\tau=0$, parameters for which the spin filtering is almost ideal. 
It almost recovers our previous result in Ref.~\cite{YU2020}. 
As seen, the lower band is spin degenerate.
The upper band can be effectively described by two decoupled ladders threaded by a fractional flux in each rung [Fig.~\ref{fig:setupDNA} (b)]~\cite{YU2020}.
The flux is induced by the intra-atomic SOI and the helical structure.
The Hamiltonians of two decoupled ladders, $\hat{\mathcal H}^{}_+$ and $\hat{\mathcal H}^{}_-$, are~\cite{YU2020},
\begin{align}
\hat{\mathcal H}^{}_\pm =&\sum_{n}^{} ( - J \hat{a}_{n+1;\pm}^\dagger \hat{a}^{}_{n;\pm} +{\mathrm H.c.}  ) \nonumber \\ & \pm  p \Delta^{}_{so} \, \hat{a}_{n;\pm}^\dagger \left[\begin{array}{cccc} 0 & e^{-i p \phi^{}_n} \\ e^{i p \phi^{}_n} & 0 \end{array} \right] \hat{a}^{}_{n;\pm} \ ,  \label{eqn:H_pm}
\end{align}
where
\begin{align}
\hat{a}^{\dagger}_{n;+} = \left[ \begin{array}{cc} \hat{a}^{\dagger}_{n;p_x \uparrow} & \hat{a}^{\dagger}_{n;p_z \downarrow} \end{array} \right],
\ \ \hat{a}^{\dagger}_{n;-} = \left[ \begin{array}{cc} \hat{a}^{\dagger}_{n;p_z \uparrow} & \hat{a}^{\dagger}_{n;p_x \downarrow} \end{array} \right] \, ,
\label{eqn:ann_FB}
\end{align}
and $\hat{a}^{}_n = e^{i L_z p \phi_n} \hat{c}^{}_n$.
The two ladders are time-reversed of one another $\hat{\Theta} \hat{\mathcal H}^{}_+ \hat{\Theta} ^{-1} = \hat{\mathcal H}^{}_-$~\cite{YU2020}, which is reminiscent of the quantum spin Hall system~\cite{Bernevig2013}. 
At the boundary of the Brillouin zone,  $k=\pm \pi$, i.e., around $E=\pm 2 J \cos(\pi/N)$ (indicated by dotted lines), there are left-going $\uparrow$($\downarrow$)-spin states and right-going $\downarrow$($\uparrow$)-spin states.
The width of the energy window in which the helical states reside is compatible with the intra-atomic SOI, $\Delta^{}_{so}$. 
These states are responsible for the spin filtering~\cite{YU2020}. 
They are degenerate and, away from the prescribed condition, the degeneracy is lifted: 
Panel (b) is drawn for $\alpha=1.2$, which also broadens the width of the bands.
Pairs of up and down spin states propagating at opposite directions are clearly observed around  $E=\pm 2 J \cos(\pi/N)$ (dotted lines).
Since we take $\tau=0$, the pitch is zero, $\Delta h=0$, and this ideal situation is realized only hypothetically.

In panel (c), we chose the parameter $\varphi \neq -\Delta \phi$. 
In this case the Hamiltonian cannot be separated into two time-reversed ones as in Eqs.~(\ref{eqn:H_pm}). 
The deviation from $\varphi = -\Delta \phi$ induces the mixing between $\sigma$  and $\pi$ orbitals. 
As seen in the figure, there appears spin splitting in the lower band, induced by the inter-atomic SOI:
It results from the intra-atomic SOI combined with the mixing of the $\sigma$ and $\pi$ orbitals on neighboring atoms due to the curved geometry~\cite{HuertasHernandoPRB2006,VarelaPRB2016}.
The inter-atomic Rashba-like SOI induced in this way is reduced by $\sim J \Delta_{so}/K_t$ as compared with the bare intra-atomic SOI.

Panel (d) shows the result for parameters taken to mimic a DNA molecule. 
The number of sites in each turn is $N=10$ corresponds to the number of base pairs. 
The dimensionless torsion is taken to be $\tau=0.48$, as estimated for a B-form DNA: $R=1 \mathrm{nm}$ and $\Delta h=3.4 \mathrm{nm}$~\cite{Sasao2019}.
The finite torsion approximately reduces the energy window of the helical states by ${\kappa} \Delta^{}_{so}$~\cite{YU2020}. 
In panel (d), one still finds helical states close to the top and bottom of the upper band.

Although our model realizes the spin current without breaking time-reversal symmetry, it is not sufficient to explain the experimentally observed magneto conductance~\cite{Xie2011,Mishra2020}. 
Earlier papers~\cite{Yang2019,Naaman2020,Yang2020,Yang2020a} argued that the Onsager relations forbid linear magneto-conductance in chiral molecules which connect a ferromagnet with a normal metal, and attributed the observed non-linear magneto-conductance to electro-electron~\cite{Fransson2019} or electron-phonon~\cite{Fransson2020,Michaeli2022} interactions. 
Here we showed that a linear spin conductance can be generated even when time-reversal symmetry is not broken. 
However, the full explanation of the experimental observations probably also need these additional interactions.

\section*{Conclusion}
\label{conclusion}

Chirality-induced spin selectivity, invoked by the spin-orbit interaction, has been discussed within the single-particle picture.
The appearance of spin current in a time-reversal symmetric system when two orbital channels participate in the transport is demonstrated. 

We analyze the helical symmetry of the infinite $p$-orbital helical atomic chain with intra-atomic spin-orbit interaction and a strong crystalline field along the helix introduced in Ref.~\cite{YU2020}. 
The helical symmetry imposes constraints on the nearest-neighbor $p$-orbital hopping matrix elements: 
They are parameterized by 3 independent real numbers. 
We explore parameters away from the condition analyzed in Ref.~\cite{YU2020}, in which the ideal spin-filtering is realized. 
We demonstrate that for a wide range of parameters, pairs of up and down spins propagating along opposite directions survive around the top and the bottom of the band. 
These helical states in the infinite atomic chain would be responsible for spin filtering in the two terminal transport experiments. 
The deviation from the ideal spin-filtering condition would not spoil our previous findings~\cite{YU2020}.

As pointed out in Ref.~\cite{Gutierrez2013}, the two orbitals can be on the same helix and thus the intra-atomic SOI is sufficient for the spin-filtering.
In our simple $p$-orbital helical atomic chain, the typical energy scale of the helical states is approximately the intra-atomic SOI times the curvature of the helix.
The intra-atomic SOI is typically larger than the inter-atomic SOI induced by the mixing between $\pi$- and $\sigma$-bands and thus would be a likely candidate for explaining the CISS effect.

\section*{Acknowledgments}

This work was supported by JSPS KAKENHI Grants No. 18KK0385, No. 20H01827 and No. 20H02562.


\begin{shaded}
\noindent\textsf{\textbf{Keywords:} \keywords}
\end{shaded}


\setlength{\bibsep}{0.0cm}
\bibliographystyle{Wiley-chemistry}
\bibliography{manu_IJC}

\begin{thebibliography}{10}

\bibitem{Goehler2011}
B.~G{\"o}shler, V.~Hamelbeck, T.~Z. Markus, M.~Kettner, G.~F. Hanne, Z.~Vager,
  R.~Naaman, H.~Zacharias, \emph{Science} \textbf{2011}, \emph{331}, 894.

\bibitem{Xie2011}
Z.~Xie, T.~Z. Markus, S.~R. Cohen, Z.~Vager, R.~Gutierrez, R.~Naaman,
  \emph{Nano Lett.} \textbf{2011}, \emph{11}, 4652.

\bibitem{Mishra2020}
S.~Mishra, A.~K. Mondal, S.~Pal, T.~K. Das, E.~Z.~B. Smolinsky, G.~Siligardi,
  R.~Naaman, \emph{J. Phys. Chem. C} \textbf{2020}, \emph{124}, 10776.

\bibitem{Naaman2019}
R.~Naaman, Y.~Paltiel, D.~H. Waldeck, \emph{Nature Reviews Chemistry}
  \textbf{2019}, \emph{3}, 1.

\bibitem{Waldeck2021}
D.~H. Waldeck, R.~Naaman, Y.~Paltiel, \emph{APL Materials} \textbf{2021},
  \emph{9}, 040902.

\bibitem{Naaman2012}
R.~Naaman, D.~H. Waldeck, \emph{J. Phys. Chem. Lett.} \textbf{2012}, \emph{3},
  2178.

\bibitem{Michaeli2016}
R.~N. K.~Michaeli, N.~Kantor-Uriel, D.~H. Waldeck, \emph{Chem. Soc. Rev.}
  \textbf{2016}, \emph{45}, 6478.

\bibitem{Michaeli2017}
K.~Michaeli, V.~Varade, R.~Naaman, D.~H. Waldeck, \emph{J. Phys.: Condens.
  Matter} \textbf{2017}, \emph{29}, 103002.

\bibitem{Evers2022}
F.~Evers, A.~Aharony, N.~Bar-Gill, O.~Entin-Wohlman, P.~Hedegard, O.~Hod,
  P.~Jelinek, G.~Kamieniarz, M.~Lemeshko, K.~Michaeli, V.~Mujica, R.~Naaman,
  Y.~Paltiel, S.~Refaely-Abramson, O.~Tal, J.~Thijssen, M.~Thoss, J.~M. van
  Ruitenbeek, L.~Venkataraman, D.~H. Waldeck, B.~Yan, L.~Kronik, \emph{The
  Journal of Physical Chemistry Letters} \textbf{2022}, \emph{13}, 7.

\bibitem{Guo2012}
A.-M. Guo, Q.-F. Sun, \emph{Phys. Rev. Lett.} \textbf{2012}, \emph{108},
  218102.

\bibitem{Gutierrez2013}
R.~Gutierrez, E.~Diaz, C.~Gaul, T.~Brumme, F.~Dominguez-Adame, G.~Cuniberti,
  \emph{J. Phys. Chem. C} \textbf{2013}, \emph{117}, 22276.

\bibitem{Guo2014}
A.-M. Guo, Q.-F. Sun, \emph{Proc. Natl. Acad. Sci. USA} \textbf{2014},
  \emph{111}, 11658.

\bibitem{Matityahu2016}
S.~Matityahu, Y.~Utsumi, A.~Aharony, O.~Entin-Wohlman, C.~A. Balseiro,
  \emph{Phys. Rev. B} \textbf{2016}, \emph{93}, 075407.

\bibitem{Michaeli2019}
K.~Michaeli, R.~Naaman, \emph{J.~Phys.~Chem.~C} \textbf{2019}, \emph{123},
  17043.

\bibitem{Geyer2020}
M.~Geyer, R.~Gutierrez, G.~Cuniberti, \emph{J. Chem. Phys.} \textbf{2020},
  \emph{152}, 214105.

\bibitem{YU2020}
Y.~Utsumi, O.~Entin-Wohlman, A.~Aharony, \emph{Phys. Rev. B} \textbf{2020},
  \emph{102}, 035445.

\bibitem{SierraBioMol2020}
M.~A. Sierra, D.~Sanchez, R.~Gutierrez, G.~Cuniberti, F.~Dominguez-Adame,
  E.~Diaz, \emph{Biomolecules} \textbf{2020}, \emph{10}, 49.

\bibitem{Liu2021}
Y.~Liu, J.~Xiao, J.~Koo, B.~Yan, \emph{Nature Materials} \textbf{2021},
  \emph{20}, 1.

\bibitem{Wolf2022}
Y.~Wolf, Y.~Liu, J.~Xiao, N.~Park, B.~Yan, \emph{arXiv:2208.00043}.

\bibitem{Michaeli2022}
D.~Klein, K.~Michaeli, \emph{arXiv:2208.02530}.

\bibitem{Bardarson2008}
J.~H. Bardarson, \emph{J. Phys. A: Math. Theor.} \textbf{2008}, \emph{41},
  405203.

\bibitem{Eto2005}
M.~Eto, T.~Hayashi, Y.~Kurotani, \emph{J. Phys. Soc. Japan} \textbf{2005},
  \emph{74}, 1934.

\bibitem{Entin-WohlmanPRB2010}
O.~Entin-Wohlman, A.~Aharony, Y.~Tokura, Y.~Avishai, \emph{Phys. Rev. B}
  \textbf{2010}, \emph{81}, 075439.

\bibitem{NagaevPRB2014}
K.~E. Nagaev, A.~S. Goremykina, \emph{Phys. Rev. B} \textbf{2014}, \emph{89},
  035436.

\bibitem{Otsuto2021}
R.~Otsuto, Y.~Yatabe, H.~Akera, \emph{Phys. Rev. B} \textbf{2021}, \emph{104},
  035431.

\bibitem{Zoellner2020}
M.~S. Z{\"o}llner, S.~Varela, E.~Medina, V.~Mujica, C.~Herrmann, \emph{J. Chem.
  Theory Comput.} \textbf{2020}, \emph{16}, 2914.

\bibitem{JJSakurai1985}
J.~J. Sakurai, \emph{Modern Quantum Mechanics}, Benjamin/Cummings, Menlo Park,
  California \textbf{1985}.

\bibitem{Bernevig2013}
B.~A. Bernevig, T.~L. Hughes, \emph{Topological Insulators and Topological
  Superconductors}, Princeton University Press \textbf{2013}.

\bibitem{Gutierrez2012}
R.~Gutierrez, R.~N. E.~Diaz, G.~Cuniberti, \emph{Phys. Rev. B} \textbf{2012},
  \emph{85}, 081404(R).

\bibitem{HuertasHernandoPRB2006}
D.~Huertas-Hernando, F.~Guinea, A.~Brataas, \emph{Phys. Rev. B} \textbf{2006},
  \emph{74}, 155426.

\bibitem{VarelaPRB2016}
S.~Varela, V.~Mujica, E.~Medina, \emph{Phys. Rev. B} \textbf{2016}, \emph{93},
  155436.

\bibitem{Sasao2019}
N.~Sasao, H.~Okada, Y.~Utsumi, O.~Entin-Wohlman, A.~Aharony, \emph{J. Phys.
  Soc. Jpn.} \textbf{2019}, \emph{88}, 064702.

\bibitem{Yang2019}
X.~Yang, C.~H. van~der Wal, B.~J. van Wees, \emph{Phys. Rev. B} \textbf{2019},
  \emph{99}, 024418.

\bibitem{Naaman2020}
R.~Naaman, D.~H. Waldeck, \emph{Phys. Rev. B} \textbf{2020}, \emph{101},
  026403.

\bibitem{Yang2020}
X.~Yang, C.~H. van~der Wal, B.~J. van Wees, \emph{Phys. Rev. B} \textbf{2020},
  \emph{101}, 026404.

\bibitem{Yang2020a}
X.~Yang, C.~H. van~der Wal, B.~J. van Wees, \emph{Nano Lett.} \textbf{2020},
  \emph{20}, 6148.

\bibitem{Fransson2019}
J.~Fransson, \emph{J. Phys. Chem. Lett.} \textbf{2019}, \emph{10}, 7126.

\bibitem{Fransson2020}
J.~Fransson, \emph{Phys. Rev. B} \textbf{2020}, \emph{102}, 235416.

\end{thebibliography}









\end{document}